# Reply to: Three papers regarding the origin of GN-z11-flash

Linhua Jiang[1,2], Shu Wang[1,2], Bing Zhang[3], Nobunari Kashikawa[4,5], Luis C. Ho[1,2], Zheng Cai[6], Eiichi Egami[7], Gregory Walth[8], Yi-Si Yang[9,10], Bin-Bin Zhang[9,10], Hai-Bin Zhao[11,12]

[1]*Kavli Institute for Astronomy and Astrophysics, Peking University, Beijing, China*

[2]*Department of Astronomy, School of Physics, Peking University, Beijing, China*

[3]*Department of Physics and Astronomy, University of Nevada, Las Vegas, NV, USA*

[4]*Department of Astronomy, Graduate School of Science, The University of Tokyo, Tokyo, Japan*

[5]*Optical and Infrared Astronomy Division, National Astronomical Observatory, Tokyo, Japan*

[6]*Department of Astronomy, Tsinghua University, Beijing, China*

[7]*Steward Observatory, University of Arizona, Tucson, AZ, USA*

[8]*Observatories of the Carnegie Institution for Science, Pasadena, CA, USA*

[9]*School of Astronomy and Space Science, Nanjing University, Nanjing, China*

[10]*Key Laboratory of Modern Astronomy and Astrophysics (Nanjing University), Ministry of Education, China*

[11]*CAS Key Laboratory of Planetary Sciences, Purple Mountain Observatory, Chinese Academy of Sciences, Nanjing, China*

[12]*CAS Center for Excellence in Comparative Planetology, Hefei, China*

In Jiang et al.[1], we detected a bright flash (hereafter GN-z11-flash) that appeared as compact continuum emission during our Keck MOSFIRE observations[2] of the galaxy GN-z11 at $z \approx 11$. We performed a comprehensive analysis of the origin of the flash using all available information and our current understanding of known man-made objects or moving objects in the solar system. We found that GN-z11-flash was likely a rest-frame UV flash associated with a long gamma-ray burst (GRB) from GN-z11. Kann et al.[3] further showed that its luminosity is consistent with the broad context of the early UV emission of GRBs. Padmanabhan & Loeb[4] proposed another explanation that the flash was likely from a shock-breakout of a Pop III supernova explosion occurring in GN-z11. Recently, Steinhardt et al.[5], Michałowski et al.[6], and Nir et al.[7] reported that GN-z11-flash was more likely from a satellite. While one cannot completely rule out the possibility of unknown satellites (or debris), we find that either the chance probabilities of being a satellite estimated by these authors have been largely overestimated or their identified satellites have been ruled out in our original analysis. Our new calculations show that the probability of GN-z11-flash being a satellite is still lower than that of it being a signal originated from GN-z11.

Steinhardt et al.[5] searched 12,300 MOSFIRE images in the archive and found 27 single-exposure transients. They defined their chance probability as the fraction of the images with transients. This is very different from our definition of the probability that one detects an event like GN-z11-flash in our target region of 0.9″ × 0.7″, or the probability that one object moves across this region after we rule out the probability of all known sources. Here 0.9" is the slit width and 0.7" is our tolerance of the position uncertainty along the slit (roughly four pixels). Let us assume a hypothetical case in which MOSFIRE has a field-of-view (FoV) of 5 deg$^2$. From a quick calculation considering the real FoV of 3′×6′ and the information provided by Steinhardt et al.[5], we would see one transient in each exposure on average (in the dawn or dusk hours; see our analysis later). Based on the definition by Steinhardt et al., the probability would be 100%. This simple fraction is clearly inappropriate. Steinhardt et al.[5] also argued that our slope of the GN-z11-flash spectrum is consistent with a reflected Solar spectrum (slope −3.5) that they identified. However, our slope has a large uncertainty due to the low signal-to-noise ratio in a short spectral baseline, and its 1$\sigma$ range is between −3.6 and −2.8 that would be consistent with a wide range of objects.

Steinhardt et al.[5] did not analyze the GN-z11-flash event itself, and their transients are not GN-z11-flash-like transients. In Jiang et al.[1], we performed a comprehensive analysis of GN-z11-flash using all available information, including observational information, slit mask design, spectral properties, and the properties of man-made satellites and moving objects in the solar system. Such information is critical. For example, the information of the observing time and site can be used to rule out satellites with height <4000 km. Steinhardt et al.[5] included all transients that they identified. We looked into the eight transients (excluding GN-z11-flash) in GOODS-N considered by Steinhardt et al.[5] This famous deep field is at high Galactic latitude without bright stars and is far from the plane of the Ecliptic. We found that six of them were observed in the dawn hours, meaning that they were low-Earth orbit satellites. Another one shows an extended spectral feature (MF.20170304.45898), suggesting an aircraft origin. There is only one transient in this sample that appears point-like and was observed during midnight. Since its two-dimensional (2D) spectrum is noisy, and it is not associated with a known object in the imaging data, we did not analyse it in detail. The transient event that Steinhardt et al.[5] particularly used (their Figures 1, 2, and 3) is toward Abell 1689 at Dec ≈ −1.5°, where one expects to see geosynchronous satellite trails (see more discussion later).

Another critical difference between GN-z11-flash and the transients identified by Steinhardt et al.[5] is that GN-z11-flash is positionally associated with the known galaxy GN-z11, while other transients are not. The spatial location of the line and continuum emission of GN-z11 that we detected in the combined 2D spectrum has a marginal offset of ~2±1 pixels (or ~1±1 pixel) from the expected position, which is not surprising for the reasons explained in Jiang et al.[2]. The position of GN-z11-flash is consistent with the position of the detected line and continuum emission within ~1±1 pixel. We used 4 pixels as our tolerance of the position uncertainty. The transients identified by Steinhardt et al.[5] are not associated with known galaxies. When a transient does not have a corresponding object in imaging data, the explanation of a moving object becomes natural. In summary, we are not convinced that Steinhardt et al.'s sample contains GN-z11-flash-like transients, and thus their analysis can neither prove nor disprove our results.

It is worth noting that one can use archival multi-object spectral data to estimate the event rate of transients like GN-z11-flash in the following three steps: (1) identify transients in the archival data; (2) identify transients that have corresponding counterparts in imaging data; (3) perform a rigorous analysis to rule out known objects and estimate the probability of GN-z11-flash-like events. One may switch the first two steps, i.e., first to identify target positions, and then search for transient signals at these positions.

Michałowski et al.[6] reported that GN-z11-flash was likely produced by a known satellite. The authors performed a careful analysis of the satellite trajectory. They also observed the satellite in 2020 and estimated that its brightness was consistent with the brightness of GN-z11-flash. We looked into our records and found that this satellite was ruled out in our original analysis. Based on our original calculations using https://www.calsky.com (which unfortunately recently has shut down; see Jiang et al.[1]), the closest separation between this satellite and GN-z11-flash was 4.4′, and its brightness was much fainter than what was needed to produce the flash. It is not clear what causes the discrepancy between our result and the result of Michałowski et al.[6]. On the other hand, this satellite would have produced a wider line. Michałowski et al.[6] showed that the angle between the slit and the trajectory was ~61°. The slit width was 0.9″, so the satellite would travel $0.9 \times \cot(61°) \approx 0.5″$ along the slit. This is nearly the same as the line width of 0.6″ shown in Extended Data Figure 1a in Jiang et al.[1], and thus we expect to see a wider line if the angle is 61°. We may estimate a probability if this angle has an uncertainty (which is currently unknown). This needs further investigation. In principle, imaging data taken at the time of GN-z11-flash for the same piece of the sky would provide direct evidence for any potential transients or satellites. Thus far we have been unable to locate such images.

Nir et al.[7] reported that the GN-z11-flash event can be a satellite glint. We cannot rule out this possibility, especially if a glint was from unknown satellites (or debris). On the other hand, the chance probability for GN-z11-flash-like events was overestimated by Nir et al.[7]. The analysis of Nir et al.[7] was based on Nir et al.[8] entitled "A high-rate foreground of sub-second flares from geosynchronous satellites", suggesting that their detections were from geosynchronous satellites. Nir et al.[7] did not observe high-Dec fields, and their high-Dec data were from Corbett et al.[9]. The telescope used by Corbett et al.[9] has a pixel size of 13.2″ (PSF ≥26″). It cannot tell whether an object is a point source. It is even impossible to tell whether a detection is a cosmic ray if it appears only once in images. The authors found that the majority of their detections are false positives. Therefore, we will use the detection rates from Nir et al.[7,8] only. The telescope used by Nir et al.[7,8] has a pixel size of 2.3″ (PSF ≥5″). It is also difficult to distinguish between GN-z11-flash-like sources (PSF ~ 0.5″-0.6″) and more extended sources. But in the following analysis we will ignore this fact and assume that the objects detected by Nir et al.[7,8] are all point sources.

We compare the GRB rate with the rate of satellite glints that would appear like GN-z11-flash. The GRB rate is about 3 sky$^{-1}$ day$^{-1}$, or $10^{-4}$ deg$^{-2}$ day$^{-1}$. We start with the satellite flare rate of about 10 deg$^{-2}$ day$^{-1}$ from Nir et al.[7]. We first estimate the fraction of high-Earth orbit satellites at high Dec (high inclination angle) using our current understanding of these satellites. The vast majority of them are geosynchronous satellite orbiting around the Earth equator. A small fraction of them are navigation systems and are usually well catalogued. Therefore, the fraction of high-Dec satellite glints must be very small. Nir et al.[8] mentioned that nearly all their repeating satellite glints were from geosynchronous satellites. Here we conservatively assume that this fraction is somewhere between 1/100 and 1/10, for example 1/30. This brings the rate of ~10 $\text{deg}^{-2}$ $\text{day}^{-1}$ down to ~0.3 $\text{deg}^{-2}$ $\text{day}^{-1}$.

We then calculate the fraction of the sky covered by galaxies. Nir et al.[7] used the galaxy number density in the Hubble Deep Field (HDF) and obtained a coverage of ~1%. HDF goes very deep and reaches ~28 mag (depending on the band). This is well beyond the limit that GRBs can be typically detected so far. We use the method by Nir et al.[7] and count galaxies down to 26 mag (similar to the brightness of GN-z11). We find a coverage fraction of ~0.1%, one order of magnitude lower than the value by Nir et al.[7]. Using this fraction, we further reduce the rate of ~10 $\text{deg}^{-2}$ $\text{day}^{-1}$ to ~$3\times10^{-4}$ $\text{deg}^{-2}$ $\text{day}^{-1}$. Next, we consider the constraint from trajectory. As explained in Jiang et al.[1], if GN-z11-flash was caused by a moving object, its trajectory should be nearly perpendicular to the slit. We can assume that the fraction is roughly 10%, which brings the rate further down to ~$0.3\times10^{-4}$ $\text{deg}^{-2}$ $\text{day}^{-1}$. This is comparable to the GRB rate of ~$10^{-4}$ $\text{deg}^{-2}$ $\text{day}^{-1}$.

There are other constraints that can potentially further reduce the above rate, including object brightness, known objects, etc. For example, one object in Nir et al.[8] produced ~20 flares on average. Figure 2 of Nir et al.[8] shows a series of flares with fast varying brightness produced by one object within a duration of a few minutes. In the above analysis, we (as well as Nir et al.[7]) included all flares, i.e., we counted the number of flares, not the number of objects. But for a given transient event with a known brightness, only 1 or 2 flares in the series match its brightness. Thus, our above calculations have overestimated this rate. We do not discuss other constraints in detail.

Finally, we emphasize that we did not conclusively claim that GN-z11-flash was a GRB flash in our original paper. We just reported this event and provided our most probable interpretation. As mentioned earlier, we cannot completely rule out the possibility of unknown satellites (or debris). Despite this fact, our new calculations have suggested that our original conclusion remains valid.

## Acknowledgements

We acknowledge support from the National Science Foundation of China (11721303, 11890693, 11991052) and the National Key R&D Program of China (2016YFA0400702, 2016YFA0400703). We thank E. O. Ofek and M. Michałowski for helpful discussions.